\documentclass[twocolumn,pre,amsmath,amssymb]{revtex4}
\usepackage{verbatim,graphicx,dcolumn}

\newcommand{\beg}{\begin{equation}}
\newcommand{\eeq}{\end{equation}}
\newcommand{\vofr}{{\bf r}}

\begin{document}

\title{Statistical Mechanics of Histories: A Cluster Monte Carlo Algorithm}

\author{Natali Gulbahce, Francis J. Alexander, Gregory Johnson}
\affiliation{Los Alamos National Laboratory, P.O.Box 1663, \\
Los Alamos, NM, 87545.}

\begin{abstract}

We present an efficient computational approach to sample the histories of 
nonlinear stochastic processes. This framework builds upon recent 
work  on casting a $d$-dimensional stochastic 
dynamical system into a $d+1$-dimensional equilibrium system using the path 
integral approach. We introduce a cluster algorithm that efficiently samples histories 
and discuss how to include measurements that are available into the estimate of the 
histories. This allows our approach to be applicable to the simulation of rare events 
and to optimal state and parameter estimation. We demonstrate the utility of this 
approach for $\phi^4$ Langevin dynamics in two spatial dimensions where our 
algorithm improves sampling efficiency up to an order of magnitude.
\end{abstract}
\maketitle

\section{\label{sec1}Introduction}
Onsager and Machlup~\cite{onsager} pioneered the path ensemble
approach to classical stochastic processes in 1953, only a few years
after Feynman's seminal work on quantum systems~\cite{feynman}.
Despite nearly simultaneous origins however, the computational
application of this framework to classical systems lagged far behind
its quantum counterpart. While Monte Carlo methods were applied to
lattice gauge theory and quantum condensed matter systems in the early
1970's~\cite{wilson}, only within the past decade has the Onsager-Machlup 
approach become practical for computational modelling of classical nonequilibrium processes.

Following the analytical results of Domany~\cite{domany} for $\mbox{2+1}$ 
dimensional Potts models, computational work began with 
Zimmer~\cite{zimmer} who devised a Monte Carlo algorithm to sample
entire space-time configurations, or histories, of a kinetic Ising model. 
This work demonstrated the utility of the Monte Carlo approach where 
histories can be conditioned on rare events. 
Olender and Elber~\cite{olender} used a similar approach to circumvent 
the time limitations of molecular dynamics simulations, specifically to find
reaction pathways when both the initial and final states are
known. See the work of Chandler et al.~\cite{chandler}, J\'{o}nsson 
et al.~\cite{nudged} and others using this methodology~\cite{vandeneijnden,alexander}.

In this paper we extend the computational work by presenting a
percolation-based cluster Monte Carlo approach to sample the statistical
mechanics of histories for nonlinear stochastic processes.  We also describe 
how to apply this method to rare event simulations and optimal estimation.  
The cluster algorithm we present improves the statistical sampling of 
histories in Monte Carlo simulations significantly. In traditional spatial 
cluster algorithms~\cite{sw}, the clusters represent statistically independent 
objects at a given time. In the $d+1$ dimensional mapping we introduce, the 
clusters can be interpreted as statistically independent objects in space-time.

Our goal is to determine the conditional statistics of histories for a stochastic
dynamical system ${\bf x}(t)$, given a model and incomplete
information about that system. The state vector ${\bf{x}}$ satisfies the 
following It{\^ o} process:
\begin{eqnarray}
\label{eq:stoch}
 d{\bf x}(t) &=& {\bf f}({\bf x},t)dt + (2{\bf D}({\bf x},t))^{1/2}d{\bf W}(t),
\end{eqnarray} 
where ${\bf f}({\bf x},t)$ is the force term, and the stochastic 
effects are provided  by the last term, in which the diffusion
matrix ${\bf {D}}$ acts on a vector-valued Wiener process, ${\bf {W}}$. 
We assume that the noise errors are uncorrelated, and that the 
initial value ${\bf x}(t_0)$ is a random variable with a known distribution. This 
system could represent the configuration of a protein evolving under
Brownian dynamics~\cite{olender}, the concentration of interacting 
metabolites, the locations of atoms in a crystal undergoing a structural phase
transition or nucleation, or the state of a queue in a stochastic fluid
model. The final state can also be a rare event on which the history
is conditioned. For instance, the configuration of an unfolded protein chain can be
conditioned in the initial state and the folded protein in the final state.

The probability of the dynamics generating a given history is simply
related to the probability that it experiences a certain noise history,
${\bf\eta} (t_k)  \equiv {\bf W}(t_{k+1})-{\bf W}(t_k)$, at times $t_k$, 
where $k=0,1,\ldots,T$. We incorporate this probability into the discretized
form of Eq. \eqref{eq:stoch}. In the interest of simplicity, we use the explicit 
Euler-Maruyama discretization scheme. This leads to the following:
\begin{equation}\label{eq:disc}
{\bf x}_{k+1} = {\bf x}_k + {\bf f}({\bf x}_k, t_k) \Delta t +  (2 {\bf D}({\bf x}_k, t_k))^{1/2} {\bf \eta}(t_k).
\end{equation} 

For Gaussian uncorrelated white noise with variance 
$\langle \eta \, \eta' \rangle= \delta (t-t') $, the probability distribution of noise is 
$P \{ \eta (t) \} \propto \exp (-\frac{1}{2}\sum_k |\eta(t_k)|^2/ \Delta t)$. The 
probability of a specific history is given by, $P\{\eta (t)\} \propto \exp (-S)$
where $S$ is the action of the $d$-dimensional system (equivalent to 
the Hamiltonian of the $d+1$-dimensional system). By rearranging 
terms in Eq.~\ref{eq:disc}, the form of the action becomes
\begin{eqnarray}
S&\equiv & \sum_{k = 0}^{T - 1} \frac{1}{4\Delta t} 
\Big \{ \big[{\bf x}_{k+1}-{\bf x}_k -{\bf f}({\bf
 x}_k,t_k)\Delta t \big]^{\top}   {{\bf D}({\bf x}_k,t_k)}^{-1} \nonumber \\ 
& & \big[{\bf x}_{k+1}-{\bf x}_k - {\bf f}({\bf x}_k,t_k) \Delta t \big] \Big \} \label{eq:S} 
\end{eqnarray}
where $\top$ indicates the transpose. With action $S$, the statistics of histories 
of the time dependent, stochastic dynamical system has been cast as an 
equilibrium statistical mechanical system.

Now let us incorporate the information about the system into the action functional.
For simplicity, we will assume that the information comes at discrete times $t_m$ 
where $m$ labels each observation $m=1,\ldots, M$.  These observations (e.g. 
experimental measurements) are given in a function, ${\bf{h}}({\bf {x}},t)$, and 
it is assumed to have errors denoted here by $\epsilon({\bf x},t)$, i.e.,
\beg
{\bf y }({\bf x}_m,t_m) = {\bf h}({\bf x}_m,t_m) + {\bf\epsilon}_m,\nonumber
\eeq
with error covariance, $\langle \epsilon \epsilon'\rangle ={\bf R}_m$. By using 
Bayes' rule~\cite{alexander}, the action arising from measurements becomes
\beg
\label{eq:S_m}
S_M=\sum_{m=1}^M ({\bf h}_m -{\bf y}_m)^{\top} \,{\bf R}_m^{-1}\,({\bf h}_m-{\bf y}_m).
\eeq
The action-functional,  $S_{\rm total}=S +S_M$, assigns weights 
to individual histories.  In the absence of additional information, histories unlikely to arise 
from the dynamics are given a lower weight than histories which are more likely. However, 
when there are measurements, histories which are far from the measurements are 
given lower weight than those closer to the measurements. 

\section{\label{sec2}A space-time cluster algorithm}
To sample the distribution of histories and hence to assign weights to them, various
methods have been applied (including local Monte Carlo, unigrid and
generalized hybrid Monte Carlo~\cite{alexander}). Here we describe 
a space-time cluster algorithm which is an extension of the embedded dynamics
algorithm introduced by Brower and Tamayo (BT)~\cite{brower}. Cluster algorithms 
are widely used in physics, statistics and computer science~\cite{barbu}. The 
first of these was introduced by Swendsen and Wang (SW)~\cite{sw} which is 
based on a mapping between the Potts model and a percolation problem~\cite{fortuin}. 

Brower and Tamayo extended the SW algorithm to a continuous 
field theory by embedding discrete variables (spins) into the continuous 
field in an equilibrium classical $\phi^4$ model~\cite{brower}. The 
$\phi^4$ potential is a symmetric double well potential of the form:
\begin{equation}
\label{eq:pot}
V(\vofr,t) = (a/4)\phi^4(\vofr,t) - (b/2) \phi^2(\vofr,t) 
\end{equation}
The discrete spin variables, $s_r$, label the two wells in $\phi^4$
potential such that $\phi_r = s_r |\phi_r|$. At fixed values of $|\phi(\vofr)|$ 
a ferromagnetic Ising model is embedded into the $\phi^4$ field theory which allows 
the use of the SW dynamics. The detailed procedure of the embedded dynamics is as follows:\\
$\bullet$ Update $\phi_r$ via a standard local Monte Carlo algorithm.\\
$\bullet$ Form percolation clusters dictated by the bond probability,
\beg 
p_ {rr'}= 1-e^{-\beta_{r r'}(1+s_r s_r')} = 1 - e^{-(|\phi_r||\phi_r'|+\phi_r \phi_r')},
\nonumber
\eeq
where the effective spin-spin coupling is $\beta_{r r'} =|\phi_r \phi_r'|$.
Note that $p_{rr'}$ reduces to $1-\exp(-2\beta_ {r r'})$ when the spins are the 
same sign.\\
$\bullet$ Update the Ising variables by flipping the percolation clusters
independently with probability $1/2$. If the move is accepted, flip the sign of the fields
in the cluster.

To extend the embedded dynamics to space-time, we need to redefine the clusters 
based on the discretized dynamical equation and the corresponding action as in 
Eqs.~\ref{eq:disc} and~\ref{eq:S}.  Next we illustrate this formalism with the 
$\phi^4$ field theory in $(2 + 1)$ dimensions.

We consider the discretized Langevin equation,
\begin{eqnarray}
\label{eq:phi}
\phi(\vofr,t+\Delta t) =\phi(\vofr,t) + \frac{\Delta t}{\Delta x^2}\,\Big[\sum_i \phi(\vofr_i,t)-4 \phi(\vofr,t)\Big] \nonumber\\
 +\Delta t \!\big[\!-\!a\phi(\vofr,t)^3+b \phi(\vofr,t)\big] + \sqrt \Delta t \,\eta(\vofr,t),
\end{eqnarray}
where the force term is the derivative of Eq.~\ref{eq:pot} with respect to $\phi$, and $\sum_i$ is 
sum over the nearest neighbors of $\phi(\vofr,t)$. The noise variables $\eta(\vofr,t)$ are chosen to 
be Gaussian distributed, independent random variables of mean zero and with correlations
 $ \langle \eta(\vofr, t)\eta(\vofr ', t') \rangle = 2D\delta_{\vofr, \vofr'}\delta(t-t')$. For this model 
the action becomes
\begin{eqnarray}
S &\equiv& \frac{1}{4 D \Delta t}\sum_{r,t}\Big(\phi(\vofr,t+\Delta t)-\phi(\vofr,t) -  
\Delta t \Big[\!-a\phi^3(\vofr,t)\nonumber\\
&+&b \phi(\vofr,t)\Big]-\Delta t \Big[\sum_i \phi(\vofr_i,t) -4 \phi(\vofr,t)\Big] \Big) ^2 \label{eq:S2}.
\end{eqnarray}
By expanding the square in the right side of Eq.~\ref{eq:S2}, we obtain
many cross terms representing different couplings between neighbors both in space 
and time. All of the interactions between a site and its neighbors in space and time are 
shown explicitly by Zimmer~\cite{zimmer}. Excluding the local terms (e.g. $\phi(\vofr,t)^2$), 
the interactions yielding different spin-spin couplings can be grouped into four types (using 
$(\vofr_j,t_k)$ as the reference site):\\
1.~Nearest neighbors of $(\vofr_j,t_{k-1})$  coupled to $(\vofr_j,t_k)$:
\begin{equation}
\beta_1 = 2 \Delta t \Big(\sum_i \phi(\vofr_i,t_{k-1})\Big)\,\,\phi(\vofr_j,t_k).\nonumber\\
\end{equation}
2.~Site $(\vofr_j,t_{k-1})$ coupled to $(\vofr_j,t_k)$:
\begin{equation}
\beta_2 = \Big[(2b-8)\Delta t-2a\Delta t\, \phi^2(\vofr_j,t_{k-1}) +2\Big]\phi(\vofr_j,t_k)\phi(\vofr_j,t_{k-1}).\nonumber\\
\end{equation}
3.~Nearest neighbors of $(\vofr_j,t_k)$ coupled to each other:
\begin{equation}
\beta_3 = -\Delta t^2 \Big(\sum_i \phi(\vofr_i,t_k)\Big)\Big(\sum_i \phi(\vofr_i,t_k)\Big).\nonumber\\
\end{equation}
4.~Nearest neighbors of $(\vofr_j,t_k)$ coupled to $(\vofr_j,t_k)$:
\begin{eqnarray}
\beta_4 = \Big(\sum_i \phi(\vofr_i,t_k)\Big) \Big([(8-2b)\Delta t^2  -2 \Delta t] \phi(\vofr_j,t_k) \nonumber\\
      +2a\Delta t ^2 \phi^3(\vofr_j,t_k)\Big). \nonumber
\end{eqnarray}
The probability of a site having a bond with any of its neighbors is
\begin{equation}
P_i = 1-e^{-2 \beta_i / (4D\Delta t)},
\end{equation}
where $i=1,\ldots,4$. A significant difference from BT is that the sign of $\beta_i$ is not known 
{\it a priori}. Depending on the value of $\phi$, the interaction can be either ferromagnetic 
or antiferromagnetic~\cite{antiferro}. At each step we determine whether the coupling term is 
ferromagnetic ($\beta_i>0$) or antiferromagnetic ($\beta_i<0$) and require the signs 
of spins to be the same or opposite respectively for a bond to exist. Once the clusters are defined, we 
use the same steps as BT described earlier in text.

Next, we compare the performance of this cluster method to two other algorithms, local Monte Carlo 
and unigrid~\cite{alexander}. To quantify performance, we measured the correlation time of a quantity 
$M =|\sum \phi| $, the sum of fields at all space and time points.  This quantity is analogous to the 
magnetization of a spin system. Because $M$ is a global quantity, it is one of the slowest  modes of 
the system~\cite{toral}. We remind the reader that our cluster algorithm updates the fields by changing 
the sign of fields in a flipped cluster. Therefore, by taking the absolute value of the fields we are left with the 
true correlations.  The correlation time, $\tau$, is obtained by fitting $\exp(-t/\tau)$ to the autocorrelation 
function defined as $\langle M_{t_0+t}M_{t_0}\rangle$.
\begin{table}[ht]
\caption{\label{tab:res1} Correlation times of the
magnetization $M$ for local and cluster algorithms for several noise
strengths, $D$. The system dimensions are $L=10$ and $T=100$, the
acceptance ratio, $a \approx 0.5$, $\Delta t=0.05$ and $\Delta x=1.0$. The 
length of the run was 100,000 MCS, and the data analyzed for the last 
80,000 MCS. The cluster algorithm is fastest at $D\approx 25$. }
\begin{center}
\begin{tabular}{rcc}\hline
$D$ & $\tau_{local}$ & $\tau_{cluster}$ \\\hline
1  & 947 & 775 \\
5  & 180 & 134 \\
15 & 25 &  8.8 \\
20 & 19 & 2.9\\
25 & 12 & 1.4\\
30 & 9 & 1.1\\\hline
\end{tabular}
\end{center}
\end{table}

\begin{figure}[ht]
\scalebox{0.6}{\includegraphics{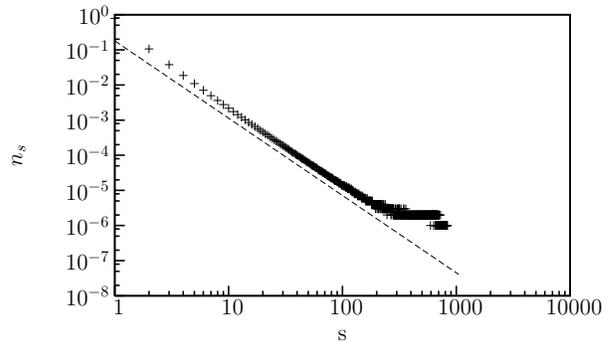}}
\caption{\label{fig:ns} The cluster size distribution at noise $D=25$. The size distribution
scales as $n_s \sim s^{-2.2}$.}
\end{figure}

The performance of the cluster algorithm depends on several factors. For a fair comparison of our 
algorithm to the local one, we used an acceptance ratio of $a\approx 0.5$ for which the local 
algorithm is empirically most efficient. The correlation times are highly dependent on noise 
strength (proportional to the square of the temperature) as is expected from 
any algorithm. We measure $\tau$'s at different noise strengths for a system of spatial dimension, 
$L=10$ with periodic boundary conditions and time dimension, $T=100$ with open boundary 
conditions. In Table~\ref{tab:res1}, these times are shown for the local and cluster algorithms 
as characterized by the decay of $C_M(t)$. The cluster algorithm performs only slightly better 
than the local algorithm at low noise strengths, and it is most beneficial at $D\approx 25$ with 
nine times more efficiency. At this noise strength, the cluster size distribution scales as 
$n_s \sim s^{-2.2}$ as shown in Fig.~\ref{fig:ns}.

We also compared the performance of the cluster algorithm to a unigrid algorithm~\cite{alexander} 
which has been shown to speed up the dynamics significantly. In Table~\ref{tab:res2}, we show 
the performance of the local, cluster and unigrid algorithms at the same noise strength ($D=25$) 
for different system sizes. The cluster algorithm correlation times are much smaller than the 
local $\tau$'s and comparable to the unigrid algorithm. 
\begin{table}[ht]
\caption{\label{tab:res2} Correlation time, $\tau$, of magnetization, $M$, with the 
local, cluster and unigrid algorithms for different system sizes, $L$ and $T$ with $D=25$. The 
cluster algorithm is a factor of nine faster than the local one, and comparable to unigrid.}
\begin{center}
\begin{tabular}{rrccc}\hline
$L$  & $T$ & $\tau_{\rm local}$ & $\tau_{\rm cluster}$ & $\tau_{\rm unigrid}$ \\\hline
8  & 32  &12.2 & 1.74  & 1.54\\
16 & 128 &11.1 & 1.50 & 1.80\\
32 & 512 &13.3 & 1.79 & 1.62\\\hline
\end{tabular}
\end{center}
\end{table}

\section{Measurements}\label{meas}
Thus far we have not included any measurements or local fields in the system.
In forecasting complex systems (e.g. weather) it is crucial to make use of data available 
to predict the path of the system. The cluster algorithm we have introduced is especially useful 
where some measurements are available. As illustrated in Eq.~\ref{eq:S}, the action 
corresponding to the measurements, $S_M$, can be added to the action, $S$, in Eq.~\ref{eq:S2}:
\begin{equation}
\label{eq:Sm2}
S_M = \sum_m \frac{[\phi(\vofr_m,t_m) - \phi_m(\vofr_m,t_m)]^2}{2\sigma_m^2},
\end{equation}
where $\phi_m$ is the value of $\phi$ measured at $\vofr_m, t_m$ with error 
variance $\sigma_m^2$. The cluster algorithm can be easily modified to incorporate the 
measurements.  The spin-spin couplings defined earlier remain the same because the measurements 
are added to the action separately and are independent of the dynamics. However 
the cluster flipping probability must be adjusted since it costs more/less to flip the sign 
of a spin if there is a measurement at that point. The local field at a site is analogous 
to having a measurement in our case. Dotsenko et al.~\cite{dotsenko} have 
discussed the probability of flipping a site in an Ising model when there are local fields 
at that site. In the presence of external field $h$, the probability of flipping a cluster
gets weighted by the local fields, i.e.,
\beg
p_{\,\rm flip} = \exp(\pm \sum_j h_j)/ [\exp(\sum_j h_j) + \exp(-\sum_jh_j) ],
\eeq
which reduces to $p_{\rm flip}=1/2$ as expected for $h=0$.

\begin{figure}[ht]
\scalebox{0.7}{\includegraphics{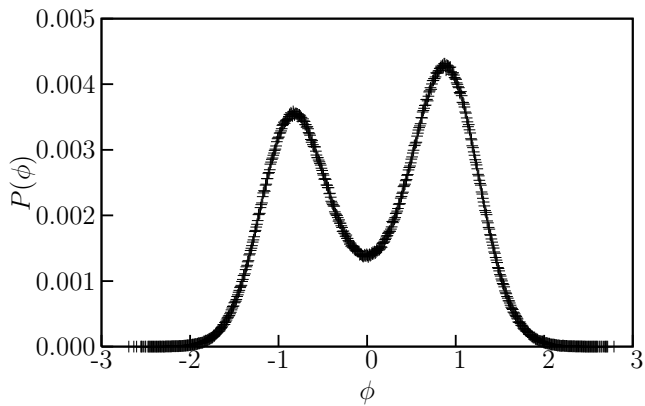}}
\caption{\label{fig:dist} Probability distribution function pf $\phi$ obtained by the cluster algorithm. The system is
$L=10$, $T=20$, and $D=1$ with measurement points, $\phi_m$, placed at every space point at every 
three time slices, $\phi_m(t<T/2)=1$ and $\phi_m(t>T/2)=-1$ with a standard deviation of $\sigma=0.02$. The 
system is driven to the negative well forced by the measurements.}
\end{figure}

Let us now derive the probability of flipping a cluster in the presence of measurements. Expanding 
the square in the right hand side of the action in Eq.~\ref{eq:Sm2} yields only 
one coupled term, $-2\phi(\vofr_m,t_m) \phi_m(\vofr_m,t_m)$. With this coupled term, the flipping 
probability becomes
\begin{equation}
p_{\rm flip} = \frac{e^{\sum_m -2 \phi(\vofr,t)\phi_m(\vofr,t)}}{e^{\sum_m 2 |\phi(\vofr,t)|\phi_m(\vofr,t)}+
e^{\sum_m -2 |\phi(\vofr,t)|\phi_m(\vofr,t)}}.
\end{equation}
We set artificial measurement points such that the system is initially in the positive well (at $t=0$), and 
it transitions into the negative well forced by the measurements. We measured the probability 
distribution function (pdf) of $\phi$ using the cluster algorithm as shown in Fig.~\ref{fig:dist}. The 
pdf obtained using the local algorithm agrees with this pdf as expected. In Table~\ref{tab:res3}, we 
show the performance of both algorithms for different system sizes ($D=25$) with four measurements 
points of variance $\sigma^2=0.01$.  The cluster algorithm consistently outperforms the local algorithm 
in the presence of the measurements.

\begin{table}[ht]
\caption{\label{tab:res3} Correlation times of $M$ for local and cluster algorithms with measurements 
at different system sizes at noise strength, $D=25$. The cluster algorithm consistently outperforms the local
one.}
\begin{center}
\begin{tabular}{rrcc}\hline
$L$  & $T$ & $\tau_{local}$ & $\tau_{cluster}$ \\\hline
8  & 32  & 10.8 & 1.7\\
12 & 72  & 11.3 & 1.6\\
16 & 128 & 11.9 & 1.5\\
24 & 288 & 12.3 & 1.7\\
\hline
\end{tabular}
\end{center}
\end{table}

\section{Discussion}\label{discuss}
In this paper, we have described a cluster Monte Carlo algorithm to sample
space-time histories of a nonlinear stochastic process.  This approach can be
applied to study pathways to rare events as well as for optimal state and
parameter estimation.

At the noise strength where the cluster size distribution scales, the cluster
algorithm outperforms the local Monte Carlo updates significantly.  We have 
not observed scaling of magnetization correlation times as a function of system 
size, therefore the observed speedup is independent of the system size.
The noise strength required to observe this scaling depends on the size of the
space-time domain. For the finite (and relatively small) systems we have
studied in this paper, this noise does not correspond to the critical
temperature in the original $D$ dimensional system.

Although the efficiency of our algorithm is comparable to the unigrid
algorithm, it can be preferred over the unigrid method when the observation of
the clusters as correlated structures is of interest. The clusters are
statistically independent space-time events, and the temporal (time-axis)
extent of these objects provides an estimate of their lifetime. For instance in
nucleation process, the correlated structures in the system, e.g. droplets, signify the
fluctuations of the metastable equilibrium~\cite{klein} and it is of interest
to measure the lifetime of these droplets directly. In the future we plan to
use this method to simulate the Ginzburg-Landau equation (model A) in order to
study nucleation and find the distribution of the lifetimes ($\tau$) of
clusters to test theoretical predictions ~\cite{klein}.

Our method is applicable to more general potentials arising from
other nonlinear stochastic partial differential equations such as Cahn-Hilliard-Cook
equation which enables the study of spinodal decomposition.

\begin{acknowledgments}
We thank G. L. Eyink, W. Klein, J. Machta, and S. K. Mitter  for 
useful discussions.  This work (LA-UR 05-8402) was funded partly by the Department of 
Energy under contracts W-7405-ENG-36 and the DOE Office of Science's ASCR 
Applied Mathematics Research program, and partly by Center for Nonlinear Studies.
\end{acknowledgments}

\end{document}